\documentclass[aps,prd,twocolumn,showpacs,amsmath,floatfix]{revtex4}
\usepackage{subfigure}
\usepackage[dvips]{graphicx}
\usepackage{mathptmx}      
\usepackage{bm}

\begin{document}

\def\be{\begin{equation}}
\def\ee{\end{equation}}
\def\bea{\begin{eqnarray}}
\def\eea{\end{eqnarray}}
\def\rra{\right\rangle}
\def\lla{\left\langle}
\def\eps{\epsilon}
\def\sgm{\Sigma^-}
\def\la{\Lambda}
\def\pv{\bm{p}}
\def\kv{\bm{k}}
\def\zv{\bm{0}}
\def\bc{B=90\;\rm MeV\!/fm^3}
\def\ms{M_\odot}

\title{Hybrid protoneutron stars within a static approach.}

\author{O. E. Nicotra}

\affiliation{Dipartimento di Fisica e Astronomia, Universit\`a di Catania and \\
INFN, Sezione di Catania Via Santa Sofia 64, 95123 Catania, Italy}

\date{\today}

\begin{abstract}
We study the hadron-quark phase transition in the interior of
protoneutron stars. For the hadronic sector, we use a microscopic equation
of state involving nucleons and hyperons derived within the
finite-temperature Brueckner-Bethe-Goldstone many-body theory, with
realistic two-body and three-body forces. For the description of quark
matter, we employ the MIT bag model both with a constant and a
density-dependent bag parameter.
We calculate the structure of protostars within a static approach. In
particular we focus on a suitable temperature profile, suggested by
dynamical calculations, which plays a fundamental role in determining the
value of the minimum gravitational mass. The maximum mass instead depends
only upon the equation of state employed.
\end{abstract}


\maketitle

\section{Introduction}\label{sec1}
After a protoneutron star (PNS) is successfully formed
in a supernova explosion, neutrinos are temporarily trapped within
the star (Prakash et al. 1997). The subsequent evolution of the PNS
is strongly dependent on the stellar composition, which is mainly
determined by the number of trapped neutrinos, and by thermal
effects with values of temperatures up to 30-40 MeV (Burrows \&
Lattimer 1986; Pons et al. 1999). Hence, the equation of state (EOS)
of dense matter at finite temperature is crucial for studying the
macrophysical evolution of protoneutron stars.\\
The dynamical transformation of a PNS into a NS 
could be strongly influenced by a phase transition to quark matter in 
the central region of the star.
Calculations of PNS structure, based on a microscopic nucleonic
equation of state (EOS), indicate that for the heaviest PNS, close
to the maximum mass (about two solar masses), the central particle density
reaches values larger than $1/\rm fm^{3}$.
In this density range the nucleon cores (dimension $\approx 0.5\;\rm fm$)
start to touch each other, and it is likely that quark 
degrees of freedom will play a role.\\
In this work we will focus on a possible hadron-quark phase transition. 
In fact, as in the case of cold NS, 
the addition of hyperons demands for the inclusion of quark 
degrees of freedom in order to obtain a maximum mass larger than the 
observational lower limit. 
For this purpose we use the Brueckner-Bethe-Goldstone (BBG) theory of nuclear 
matter, extended to finite temperature, for describing the hadronic phase and 
the MIT bag model at finite temperature for the quark matter (QM) phase. 
We employ both a constant and a density-dependent bag parameter $B$. 
We find that the presence of QM 
increases the value of the maximum mass of a PNS, and 
stabilizes it at about 1.5--1.6 $\ms$, no matter the value of the 
temperature.\\ The paper is organized as follows. In section \ref{s:qm} we 
present a new static model for PNS. Section \ref{s2:qm} is devoted to the 
description of the hadron-quark phase transition within the EOS mentioned 
above. In section \ref{s3:qm} we present the results about the structure of 
hybrid PNS and, finally, we draw our conclusions.
\section{A static model for PNS}
\label{s:qm}
Calculations of static models of protoneutron stars should be 
considered as a first step to describe these objects. In principle the 
temperature profile has to be determined via dynamical calculations 
taking into account neutrino transport properly \cite{burr}. Many 
static approaches have been developed in the past decade 
\cite{pra}\cite{strob}, 
implementing several finite temperature EOS and assuming an 
isentropic or an isothermal \cite{hean} profile throughout the star.\\
In our model we assume that a PNS in its early stage is composed of a hot, 
neutrino opaque, and 
isothermal core separated from an outer cold crust by an isentropic, 
neutrino-free intermediate layer, which will be called the envelope 
throughout the paper. 
\subsection{Isothermal core}
For a PNS core in which the strongly interacting particles are only baryons, 
its composition is determined by requirements of charge neutrality 
and equilibrium under weak semileptonic processes,
%
  $B_{1}\rightarrow B_{2}+l+\bar{\nu}_{l}$ 
and $B_{2}+l\rightarrow 
  B_{1}+\nu_{l}$, 
%
where $B_{1}$ and $B_{2}$ are baryons and $l$ is a lepton (either an 
electron or a muon). Under the condition that neutrinos are trapped in 
the system, the beta equilibrium equations read explicitly
\begin{equation}\label{betaeq}
 \mu_i = b_i \mu_n - q_i( \mu_l - \mu_{\nu_l}) \:,
\end{equation}
where $b_i$ is the baryon number, and $q_i$ the electric charge
of the species $i$.
Because of trapping, the numbers of leptons per baryon of each 
flavour ($l=e,\mu$),
$Y_{l}=x_{l}-x_{\bar{l}}+x_{\nu_{l}}-x_{\bar{\nu}_{l}}$, are conserved. 
Gravitational collapse calculations of the iron core of massive 
stars indicate that, at the onset of trapping, the electron lepton 
number is $Y_{e}\simeq 0.4$; since no muons are present at this stage we 
can impose also $Y_{\mu}= 0$. For neutrino free matter we just set 
$\mu_{\nu_{l}}=0$ in Eq.(\ref{betaeq}) and neglect the above constraints 
on lepton numbers.\\ We assume a constant value of temperature 
throughout the core and perform some calculations for a value of temperature 
ranging from 0 to 50 MeV, with and without neutrinos. The EOS employed 
is that of the BHF approach at finite temperature for the hadron phase and 
that of the MIT bag model for QM. Many more details on the nuclear matter EOS 
employed together with plots for the chemical 
composition and pressure at increasing density and temperature can be 
found in \cite{aap}.\\
\subsection{Isentropic envelope}\label{isen}
The condition of isothermality adopted for the core cannot be 
extended to the outer part of the star. Dynamical calculations 
suggest that the temperature drops rapidly to zero at the surface of the 
star; this is due to the fast cooling of the outer part of the PNS where 
the stellar matter is transparent to neutrinos. Moreover, in the early stage, 
the outer part of a PNS is characterized by a high value of the entropy 
per baryon ranging from $6$ to $10$ in units of Boltzmann's constant 
\cite{burr}.\\
For a low enough core temperature ($T\leq10$ MeV) in \cite{aap} a 
temperature profile in the shape of a step function was assumed, 
joining the hadronic EoS (BHF) with the BPS \cite{bps} plus FMT 
\cite{fmt} EoS for the cold crust. 
When the core temperature $T_{core}$ is greater than $10$ MeV, 
we consider an isentropic envelope in the 
range of baryon density from $10^{-6}$ fm$^{-3}$ to $0.01$ fm$^{-3}$ 
based on the EoS of LS \cite{ls} with the incompressibility modulus 
of symmetric nuclear matter $K=220$ MeV (LS220). Within this EoS, fixing 
the entropy per baryon $s$ to the above values ($6,8,10$) and imposing 
beta-equilibrium (neutrino-free regime) we get temperature profiles which 
rise quikly from $0$ (at $10^{-6}$ fm$^{-3}$) to values of temperature 
typical of the hot interior of a PNS (respectively 
$T_{core}=30,40,50$ MeV) at a baryon density of about $0.01$ fm$^{-3}$. 
This explicitly suggests a natural correspondance 
between the entropy of the envelope $s_{env}$ and $T_{core}$ \cite{nicot1}.\\ 
Since the energy of neutrinos, emerging from the interior, possesses some 
spreading \cite{bethe} and their transport properties vary quite a lot during 
the PNS evolution \cite{janka}, we do not think that the location of the 
neutrino-sphere, affected by the same spreading, 
is a good criterion to fix the matching density between core and 
envelope. 
All this instead indicates the possibility to 
have a blurred region inside the star where we are free to choose 
the matching density. 
To build up our model of PNS, a fine tuning of $s_{env}$  
is performed in order to have an exact 
matching between core and envelope of all the thermodynamic 
quantities (energy density, free energy density, and temperature). In 
this sense we can consider this static description of PNS with only one free 
parameter: the temperature of the core. Once $T_{core}$ is chosen, $s_{env}$ 
is fixed by matching conditions (see \cite{nicot1} for more details and results on pure baryonic PNS).
%
%
\begin{figure*}[t] 
\centering
\includegraphics[width=0.6\textwidth,clip]{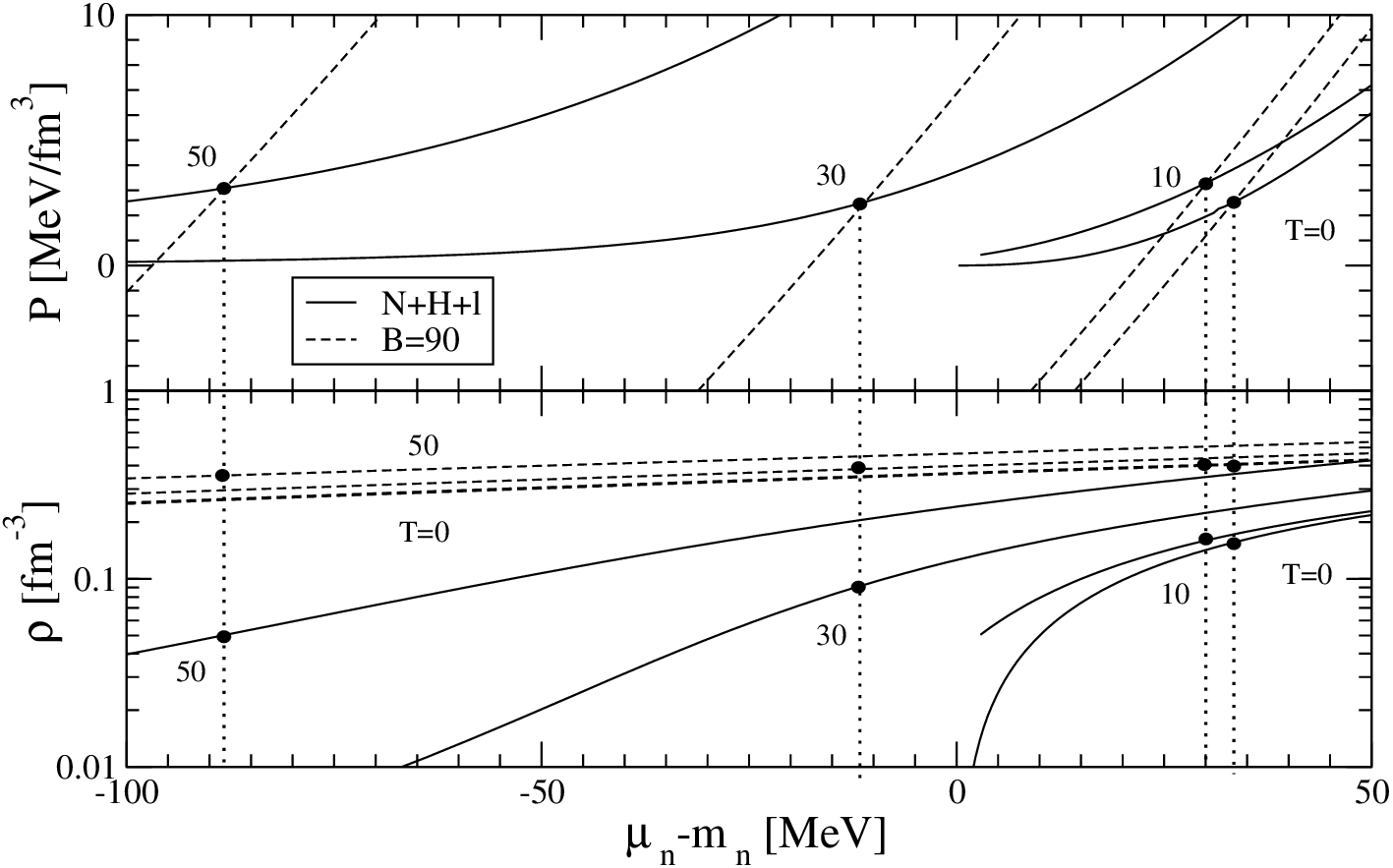}
\caption{ 
Baryon density (lower panel) and pressure (upper panel)
as a function of baryon chemical potential of beta-stable baryonic
matter (solid curves) and quark matter (dashed curves) for the neutrino-free 
case at different
temperatures $T=0,10,30,50\;\rm MeV$. The vertical dotted lines
indicate the positions of the phase transitions. A bag constant
$\bc$ is used for QM.} 
\label{f:pc}
\end{figure*} 
\section{Hadron-quark phase transition}
\label{s2:qm}
We review briefly the description of the bulk properties of
uniform QM at finite temperature, deconfined from the
beta-stable hadronic matter discussed in the previous section, by
using the MIT bag model \cite{chodos}.
In its simplest form, the
quarks are considered to be free inside a bag and the thermodynamic
properties are derived from the Fermi gas model,
where the quark $q=u,d,s$ baryon density and 
the energy density,
are given by
\bea
 \rho_q = \frac{g}{3} \int\!\!\frac{d^3k}{(2\pi)^3}\left[ f^+_q(k)-f^-_q(k) \right] \: 
\label{e:rhoq}
\\
\eps_Q = g \sum_q \int\!\!\frac{d^3k}{(2\pi)^3}\left[ f^+_q(k)+f^-_q(k) \right] E_q(k) + B \: 
\label{e:epsq}
\eea
where $g=6$ is the quark degeneracy,
$E_q(k) = \sqrt{m_q^2+k^2}$, $B$ is the bag constant 
and $f^\pm_q(k)$ are the Fermi distribution functions for the quarks and
anti-quarks. 
We have used massless $u$ and $d$ quarks, and $m_s=150$ MeV. 
It has been found \cite{alford,mit} that within the MIT bag model
(without color superconductivity) with a density-independent bag
constant $B$, the maximum mass of a NS cannot exceed a value of
about 1.6 solar masses. 
Indeed, the maximum mass increases as the
value of $B$ decreases, but too small values of $B$ are incompatible
with a hadron-quark transition density $\rho >$ 2--3 $\rho_0$ in 
nearly symmetric nuclear matter, 
as demanded by heavy-ion collision phenomenology. 
In order to overcome these restrictions of the model, 
one can introduce a density-dependent
bag parameter $B(\rho)$, and this approach was followed in
Ref.~\cite{mit}. 
This allows one to lower the value of $B$ at large
density, providing a stiffer QM EOS and increasing the
value of the maximum mass, while at the same time still fulfilling
the condition of no phase transition below $\rho \approx 3 \rho_0$
in symmetric matter. In the following we present results based on the 
MIT model using both 
a constant value of the bag parameter, $\bc$, and a
gaussian parametrization for the density dependence,
\be
 {B(\rho)} = B_\infty + (B_0 - B_\infty)\exp\left[-\beta\Big(\frac{\rho}{\rho_0}\Big)^2 \right]
\label{eq:param} 
\ee 
with $B_\infty = 50\;\rm MeV\!/fm^{3}$, $B_0 =
400\;\rm MeV\!/fm^{3}$, and $\beta=0.17$,
see Ref.~\cite{mit}. 
The introduction of a density-dependent bag 
has to be taken into account properly for the 
computation of various thermodynamical quantities; 
in particular the quark chemical potentials $\mu_q$
and the pressure $p$ are modified as
\be
 \mu_q \rightarrow \mu_q + \frac{dB(\rho)}{ d\rho} \:,\:\:\:\:\:
\label{murho}
 p \rightarrow p + \rho \frac{dB(\rho)}{d\rho} \:.
\ee
Nevertheless, due to a cancelation of the second term in (\ref{murho}), 
occurring in relations (\ref{betaeq}) for the beta-equilibrium, 
the composition at a given total baryon density remains unaffected
by this term (and is in fact independent of $B$).
At this stage of investigation, we disregard possible dependencies 
of the bag parameter on temperature and individual quark densities. 
For a more extensive discussion of this topic, the reader is referred to 
Refs.~\cite{mit}.\\ 
The individual quark chemical potentials
are fixed by Eq.~(\ref{betaeq}) with $b_q=1/3$, 
which implies:$\:\:\mu_d = \mu_s = \mu_u + \mu_l - \mu_{\nu_l} \:$.
The charge neutrality condition and the total
baryon number conservation 
together with the constraints on the lepton number $Y_l$ 
conservation allow us to determine the composition $\rho_q(\rho)$
and then the pressure of the QM phase. In both phases the contribution of 
leptons is that of a Fermi gas. 
In the range of temperature considered here ($0\div 50$ MeV) 
thermal effects are rather weak, the presence of neutrinos instead influences 
quite strongly the composition: In this case the relative
fraction of $u$ quarks increases substantially from $33\%$ to about
$42\%$, compensating the charge of the electrons that are present at
an average percentage of $25\%$ throughout the considered range of
baryon density, whereas $d$ and $s$ quark fractions are slightly lowered (see \cite{nicot2}). \\
\begin{figure*}[t] 
\centering
\includegraphics[width=0.65\textwidth,clip]{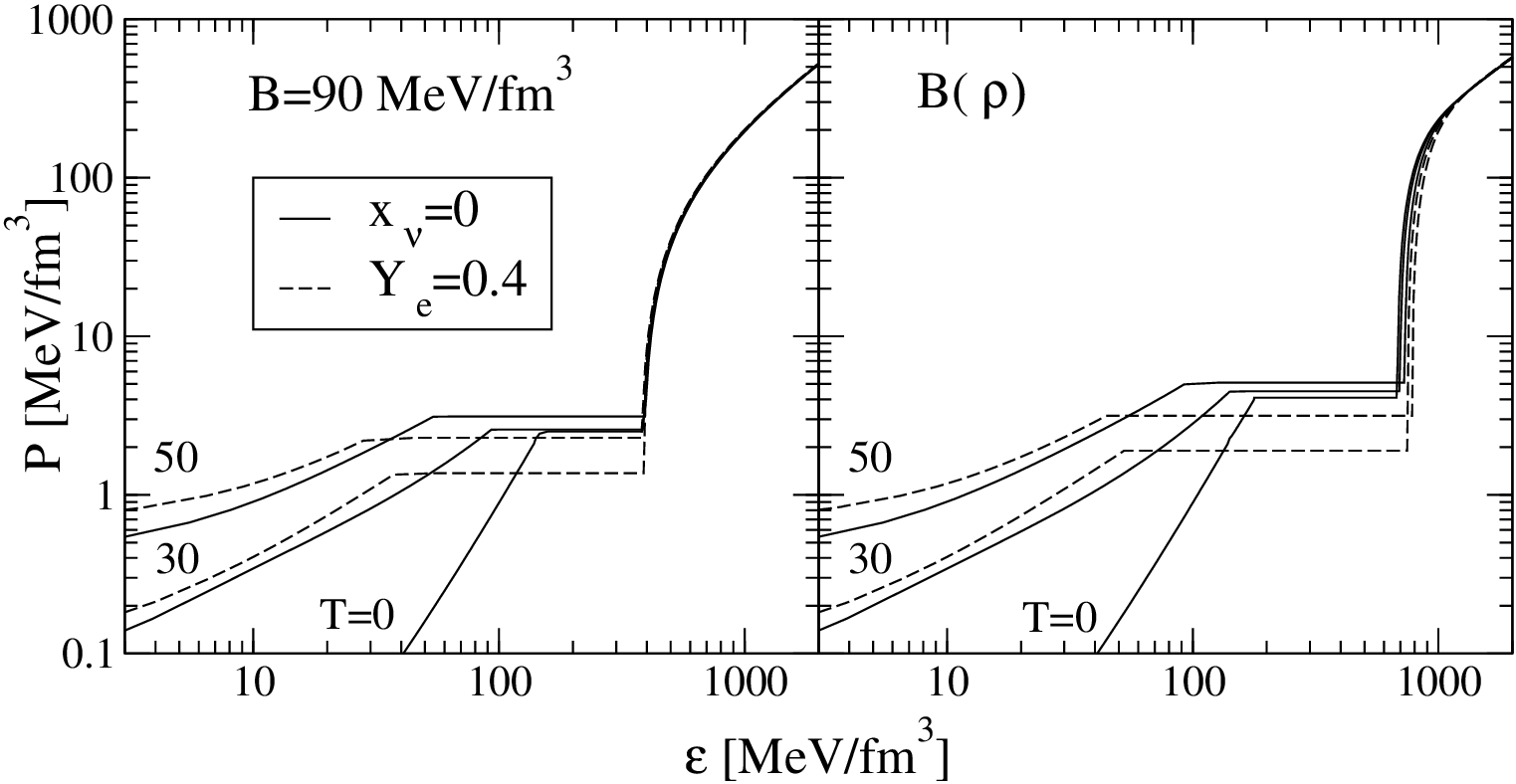}
\caption{ 
Pressure as a function of energy density for beta-stable matter
with (dashed curves) and without (solid curves) neutrino trapping 
at different temperatures $T=0$, 30, and 50 MeV
with a bag constant $\bc$ (left panel) 
or a density-dependent bag parameter (right panel).}
\label{f:pe}
\end{figure*} 
%
%
We now consider the hadron-quark phase transition in 
beta-stable matter at finite temperature.
In the present work we adopt the
simple Maxwell construction for the phase transition from the plot
of pressure versus chemical potential. 
The more general Glendenning (Gibbs) construction \cite{gle} 
is still affected by many theoretical uncertainties 
and in any case influences very little the final
mass-radius relations of massive (proto)neutron stars \cite{mit}.
We therefore display in Figs.~\ref{f:pc} the pressure
$p$ (upper panels) and baryon density $\rho$ (lower panels) as
functions of the baryon chemical potential $\mu_n$ for both
baryonic and QM phases at temperatures $T=0,10,30,50$ MeV.
The crossing points of the baryon and quark pressure curves (marked
with a dot) represent the transitions between baryon and QM
phases. The projections of these points (dotted lines) on the baryon
and quark density curves in the lower panels indicate the
corresponding transition densities from low-density baryonic matter,
$\rho_H$, to high-density QM, $\rho_Q$. \\
The main aspects of the EoS for such stellar matter are displayed in 
Fig.~\ref{f:pe} and are well summarized as follows. 
The transition density $\rho_H$ is rather low,
of the order of the nuclear matter saturation density. 
The phase transition density jump $\rho_Q-\rho_H$ is large,
several times $\rho_H$, and the model with density-dependent bag parameter 
predicts larger transition densities $\rho_H$ and larger jumps 
$\rho_Q-\rho_H$ than those with bag constant $\bc$. 
The plateaus in the Maxwell construction are thus wider for the former case. 
Thermal effects and neutrino trapping shift $\rho_H$ to lower values
of subnuclear densities and increase the density jump $\rho_Q-\rho_H$. For the 
cold case the presence of neutrinos even inhibits completely the phase 
transition \cite{nicot2}. 
\begin{figure*}[t] 
\centering
\subfigure{\includegraphics[width=12.0cm,clip]{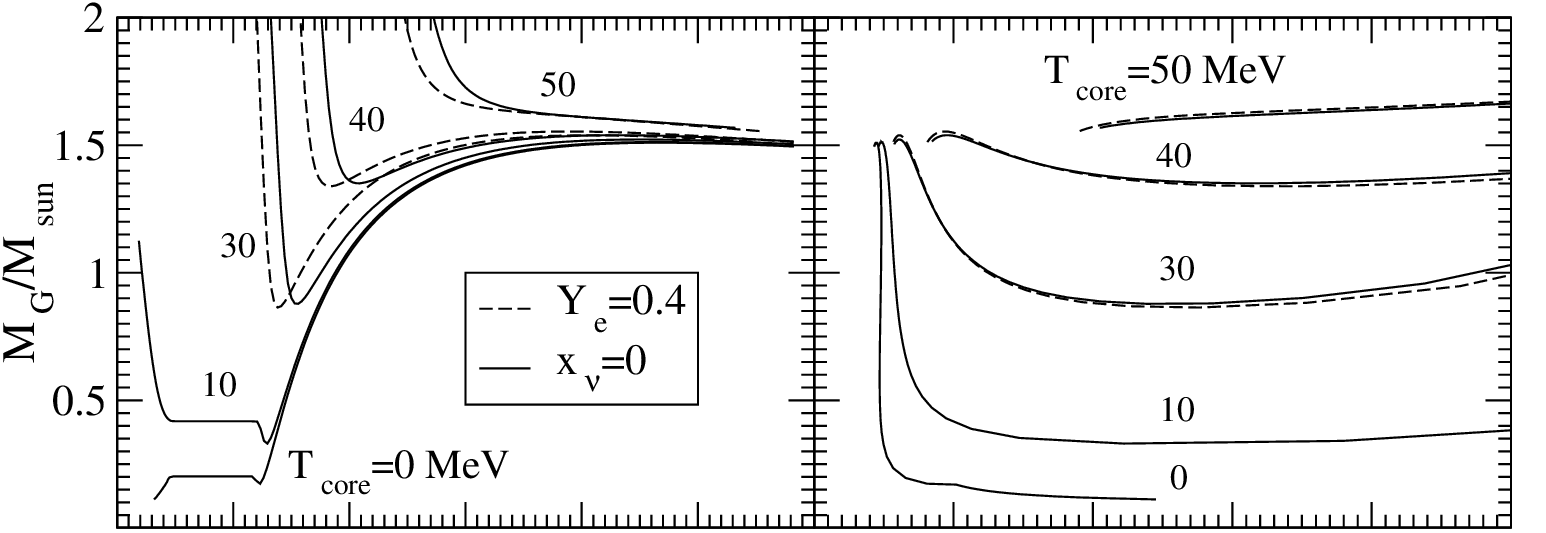}}
\\
\subfigure{\includegraphics[width=12.0cm,clip]{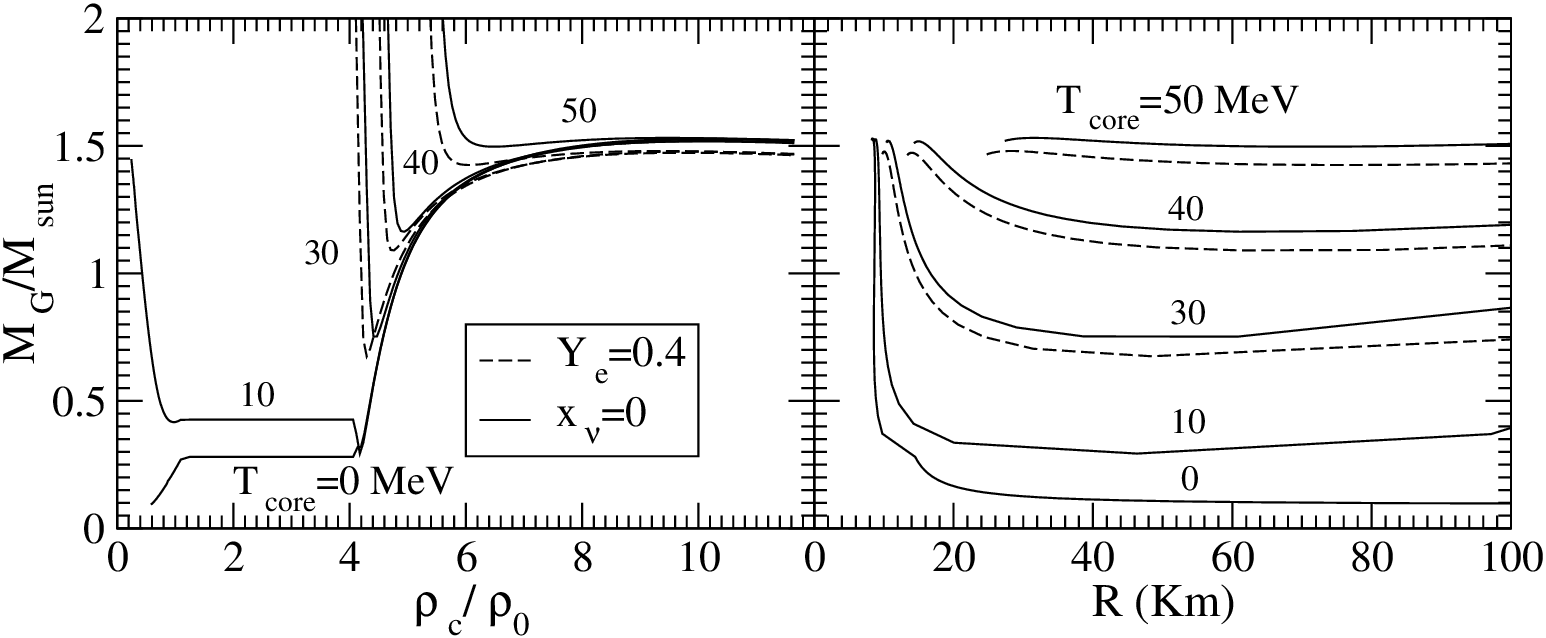}} 
\caption{(Proto)Neutron star mass-central density (left panel) and
mass-radius (right panel) relations for different core temperatures
$T=0,10,30,40,50\;\rm MeV$ and neutrino-free (solid curves) or
neutrino-trapped (dashed curves) matter. A bag constant $\bc$ is
used for QM. Same for lower panels but with a density-dependent bag
parameter.}
\label{f:mc}
\end{figure*} 
\section{Structure and stability}
\label{s3:qm}
The stable configurations of a PNS can be obtained from the 
hydrostatic equilibrium equation of Tolman, Oppenheimer, and Volkov 
\cite{shap} 
for the pressure $P$ and the gravitational mass $m$, 
once the EoS $P(\epsilon)$ is specified, being $\epsilon$ the total 
energy density. 
We schematize the entire evolution of a PNS as divided in two main 
stages. In the first, representing the early stage, the PNS is in a 
hot ($T_{core}=30\div 40$ MeV) stable configuration with a neutrino-trapped 
core and a high-entropy envelope ($s_{env}\simeq 6\div 8$). 
The second stage represents the end of the short-term cooling where the 
neutrino-free core possesses a low temperature ($T_{core}=10$ MeV) and 
the outer part can be considered as a cold crust (BPS+FMT).\\
The results are plotted in Fig.~\ref{f:mc}, where we
display the gravitational mass $M_G$ (in units of the solar mass
$\ms$) as a function of the radius $R$ (right panels) and the central
baryon density $\rho_c$ (left panels), for QM EOS with
$\bc$ and $B(\rho)$, respectively. Due to the use of the Maxwell construction, 
the curves are not continuous \cite{gle}. 
PNS in our approach are thus practically 
hybrid stars and the heaviest ones have only a thin outer layer of 
baryonic matter. 
For completeness we display the complete set of results at core
temperatures $T=0,10,30,40,50\;\rm MeV$ with and without neutrino
trapping, although only the curves with high temperatures and
neutrino trapping and low temperatures without trapping are the
physically relevant ones. We observe in any case a surprising
insensitivity of the results to the presence of neutrinos, in 
particular for the $\bc$ case, which can
be traced back to the fact that the QM EOS $p(\eps)$ in Fig.~\ref{f:pe} 
is practically insensitive to the neutrino fraction (see \cite{nicot2} for 
more details). 
On the other hand, the temperature dependence of the curves is quite
pronounced for intermediate and low-mass stars, showing a strong
increase of the minimum mass with temperature, whereas the maximum
mass remains practically constant under all possible circumstances.
Above core temperatures of about 40--50 MeV all stellar configurations
become unstable. 
Concerning the dependence on the QM EOS, we observe again
only a slight variation of the maximum PNS masses between $1.55\;\ms$
for $\bc$ and $1.48\;\ms$ for $B(\rho)$. Clearer differences exist for
the radii, which for the same mass and temperature are larger for
the $\bc$ model.\\
\section{Conclusions}
In conclusion, in this article we have extended a previous work on baryonic 
PNS \cite{aap} 
to the case of hybrid PNS. We combined the most recent microscopic baryonic 
EOS in the BHF approach involving nuclear three-body forces and hyperons
with two versions of a generalized MIT bag model for QM. The EoS employed for 
both phases are checked by phenomenological constraints. \\
We modelled the 
entropy and temperature profile of a PNS in a simplified way taking as much as 
possible care about results coming from dynamical calculations \cite{nicot1}. 
This approach allows us to study the stability of a PNS varying both the 
temperature of the core and the central density. 
\section*{Acknowledgments}
The author wishes to thank M. Baldo, G.F. Burgio, H.-J. Schulze and M. Di Toro for their kind help.

\end{document}